\documentclass[acmlarge, nonacm]{acmart}

\makeatletter
\newcommand{\myconfshort}{\acmConference@shortname}
\newcommand{\myconffull}{\acmConference@name}
\newcommand{\myconfdate}{\acmConference@date}
\newcommand{\myconfloc}{\acmConference@venue}
\AtBeginDocument{
  \fancypagestyle{firstpagestyle}{
    \fancyhead{}%
    \fancyfoot[C]{}%
  }
  \fancyhf{}
  \fancyhead[LO]{\@headfootfont\shorttitle}%
  \fancyhead[RE]{\@headfootfont\@shortauthors}%
  \fancyhead[LE]{\@headfootfont\footnotesize \myconfshort, \myconfdate, \myconfloc}%
  \fancyhead[RO]{\@headfootfont\footnotesize \myconfshort, \myconfdate, \myconfloc}%
  \fancyfoot[C]{}%
}
\makeatother
\acmBooktitle{\conffull\@ (\confshort), \confdate, \confloc}

\AtBeginDocument{%
  }

\usepackage{graphicx}
\usepackage{subcaption}
\usepackage[ruled]{algorithm2e}

\usepackage{amssymb}
\usepackage{xcolor}
\usepackage{enumitem}

\copyrightyear{2026}
\acmYear{2026}
\setcopyright{none}
\makeatletter
\def\@copyrightspace{\relax}
\makeatother
\acmConference[FAccT '26]{The 2026 ACM Conference on Fairness, Accountability, and Transparency}{June 25--28, 2026}{Montreal, QC, Canada}
\acmBooktitle{The 2026 ACM Conference on Fairness, Accountability, and Transparency (FAccT '26), June 25--28, 2026, Montreal, QC, Canada}

\begin{document}

\title{AI Washing Inflates Expected Performance but Not Interaction Outcomes: An AI Placebo Study Using Fitts' Law}

\begingroup
\renewcommand\thefootnote{}
\footnotetext{© 2026 Association for Computing Machinery. This is the author's version of the work. It is posted here for your personal use. Not for redistribution. The definitive Version of Record is published in ACM Conference on Fairness, Accountability, and Transparency (FAccT '26), https://doi.org/10.1145/3805689.3812249.}
\endgroup

\author{Nick von Felten}
\authornote{Corresponding author.}
\authornote{Both authors contributed equally to this research.}
\email{nick.vonfelten@unisg.ch}
\affiliation{%
\institution{University of St. Gallen}
\city{St. Gallen}
\country{Switzerland}
}

\author{Luisa Ella Müller}
\authornotemark[2]
\affiliation{%
  \institution{University of St. Gallen}
  \city{St. Gallen}
  \country{Switzerland}
}

\author{Johannes Schöning}
\affiliation{%
  \institution{University of St. Gallen}
  \city{St. Gallen}
  \country{Switzerland}
}

\affiliation{%
  \institution{Mohamed bin Zayed University of Artificial Intelligence}
  \city{Abu Dhabi}
  \country{United Arab Emirates}
}

\renewcommand{\shortauthors}{Nick von Felten, Luisa Müller, and Johannes Schöning}
\renewcommand{\shorttitle}{AI Washing Inflates Expected Performance: AI Placebo Study Using Fitts' Law}

\begin{abstract}
Expectations about the support of artificial intelligence (AI) may influence interaction outcomes similar to placebos. Such expectations may result from AI washing, a practice of overstating a system’s AI capabilities when actual functionality is limited. For example, some computer mice are marketed as “AI-assisted” despite lacking AI in core functions. In a within-subjects study, 28 participants completed Fitts' Law tasks with a computer mouse under three conditions: no support, supposed predictive AI support, and supposed biosignal-enhanced AI support. Objective Fitts' Law performance indicators and subjective performance expectations, perceived workload, and perceived usability were measured. Compared to baseline, participants expected significantly improved performance in placebo conditions. However, these expectations did not translate into differences in objective or subjective assessments. This paper contributes evidence that AI washing inflates user expectations without altering actual interaction outcomes, highlighting a critical transparency issue. By exposing how deceptive AI marketing can shape user expectations, we underscore the need for accountability in AI product claims. Further, we establish Fitts' Law as a rigorous methodological lens for auditing AI-labelled input devices.
\end{abstract}



\maketitle

\section{Introduction \& Related Work}
As Artificial Intelligence (AI) becomes more widely commercialised, the incentive to capitalise on its perceived capabilities introduces significant challenges for transparency and accountability. Driven by the promise of productivity gains, companies increasingly market their products as AI-powered. In some cases, however, these claims reflect only superficial integrations. For example, certain devices, such as computer mice, merely offer shortcut buttons to large language model (LLM) chatbots without altering their core functionality~\citep{logitech2024, adesso2024}. These practices exemplify \emph{AI washing}, a term referring to marketing strategies that exaggerate or misrepresent a product’s AI capabilities and that parallels greenwashing in environmental marketing~\citep{Seele2022}. In this sense, AI washing can be understood as one pathway through which so-called AI “snake oil” can operate: the commodification of inflated AI promises that causes harm when users rely on systems that fail to deliver their advertised capabilities~\citep{narayananAISnakeOil2024}. Although the term AI washing has circulated in journalism and policy discourse for several years~\citep{moore2017}, recent academic work has begun to examine its effects. For instance,~\citet{leffrang2023ai} show that labelling a standard statistical model as “AI” changes user perceptions even when it does not change advice-taking behaviour. Similar research has found that people perceive the same technologies differently depending on how they are labelled and ascribe different values to them.~\citet{langerLookItsComputer2022} found that the label "AI" leads individuals to perceive the same system as more complex and competent than labels such as "computer". In some application contexts, the "AI" label also reduced perceptions of fairness and trust. The authors note that AI labels may be used strategically to shape participant perceptions. This suggests that AI washing may do more than create unmet expectations or consumer dissatisfaction. It could also distort how users understand and evaluate a system. AI washing can thus be understood as a deceptive marketing strategy that undermines user autonomy by shaping expectations that systems may not meet. 

Extending these findings to interaction outcomes, recent work suggests that labelling a system as AI-powered may influence not only users’ initial perceptions, but also their subsequent evaluations, even when the underlying functionality remains unchanged~\citep{kosch_placebo_2022}. This makes placebo research a useful lens for examining expectancy effects of AI labels and whether these expectations translate into subjective or objective outcomes. When such outcomes arise independently of any actual technical improvement, they resemble placebo effects~\citep{shapiro_placebo_1959, hurst2020placebo, Shiv2005, kosch_placebo_2022}.~\citet[p.~199]{shapiro_placebo_1959} defines the placebo effect as \textit{“psychological, physiological or psychophysiological effect of any […] procedure given with therapeutic intent, which is independent of or minimally related to the […] specific effects of the procedure, and which operates through a psychological mechanism”}. Originally framed in medicine and psychology, placebo effects have since been observed in sports, marketing and human-computer interaction (HCI)~\citep{hurst2020placebo, Shiv2005, kosch_placebo_2022}. In HCI, placebo effects arise when users attribute improved qualities (e.g., usability or performance) to systems without any corresponding technical mechanism~\citep{kosch_placebo_2022, denisova2015placebo, bosch_illusion_2024}. Placebo research often distinguishes subjective outcomes (e.g., symptom relief) from objective measures (e.g., physiological changes). Likewise, placebo studies in HCI make similar distinctions, using subjective measures (e.g., scales) and objective indicators (e.g., error rate). In an experiment on supposed error-adaptive or physiologically adaptive AI support, subjective performance was rated higher before and after the task, while objective performance showed no difference~\citep{kosch_placebo_2022}. In games, merely mentioning AI involvement increased immersion and experience ratings~\citep{denisova2015placebo}, while narrative experiences were rated more positively when believed to be human-written~\citep{sridharan_exploring_2025}. By contrast, productivity-focused studies using validated scales such as the System Usability Scale (SUS)~\citep{brooke_sus_1996} and NASA-TLX~\citep{hart1988nasa} report no differences~\citep{kloftAIEnhancesOur2024, kosch_placebo_2022}.

The impact of AI washing on objective measures also remains inconclusive. Some studies report improvements in reaction time and speed~\citep{kloftAIEnhancesOur2024}, whereas others find no differences in error-based tasks~\citep{kosch_placebo_2022}. These mixed results could reflect reliance on unvalidated ad-hoc tasks such as word puzzles, or they may indicate that AI placebo effects are highly context-dependent~\citep{gonnermann-mullerLetsBeRealistic2025}. Furthermore, prior research has also focused primarily on software features rather than hardware devices~\citep{denisovaPowerUpsDigitalGames2019, denisova2015placebo, vaccaroIllusionControlPlacebo2018}, leaving open whether such effects extend to familiar, everyday tools.

We address these gaps by leveraging Fitts' Law, one of the most established models in HCI for evaluating input performance. It quantifies the speed--accuracy trade-off in input devices~\citep{norman_fitts_2018}. The paradigm has been widely replicated and is directly applicable to one of the most used input devices: the computer mouse~\citep{soukoreff2004}. As a ubiquitous input device that is subject to AI washing, the computer mouse provides an ideal test case for examining whether supposed AI support influences both subjective perceptions and objective performance.

In this paper, we distinguish expectancy about anticipated performance from placebo responses reflected in post-task subjective experience or objective performance improvement. Our manipulation reliably increased expectancy, but we do not observe corresponding changes in post-task self-reports or objective performance. We examine how AI-related claims shape user perceptions and behaviours beyond the systems' underlying technical properties. By exposing effects of AI washing, we contribute to a better understanding of how misleading representations are a transparency issue, that can undermine accountability and informed user consent. Such insights can support the development of more transparent communication practices and regulatory approaches that protect users from deceptive or inflated AI claims.

The contributions of this research are threefold. First, it provides controlled evidence that AI washing labels inflate users’ expectations of performance, even for a familiar input device. Second, this expectancy shift does not propagate to standard post-task self-reports (workload, usability) or objective Fitts' Law metrics (throughput, error rate, model fit), showing that AI-labelled expectations can diverge from measurable interaction outcomes. Third, it positions Fitts' Law tasks as a promising placebo-control instrument for auditing AI-labelled interaction artefacts by separating marketing-driven expectancy from measurable interaction outcomes. Overall, our findings add nuance to debates on transparency and accountability by showing how AI labels can distort user expectations even in the absence of measurable downstream effects.

\section{User Study}
\begin{figure}[t]  
\begin{center}
\includegraphics[width=\linewidth]{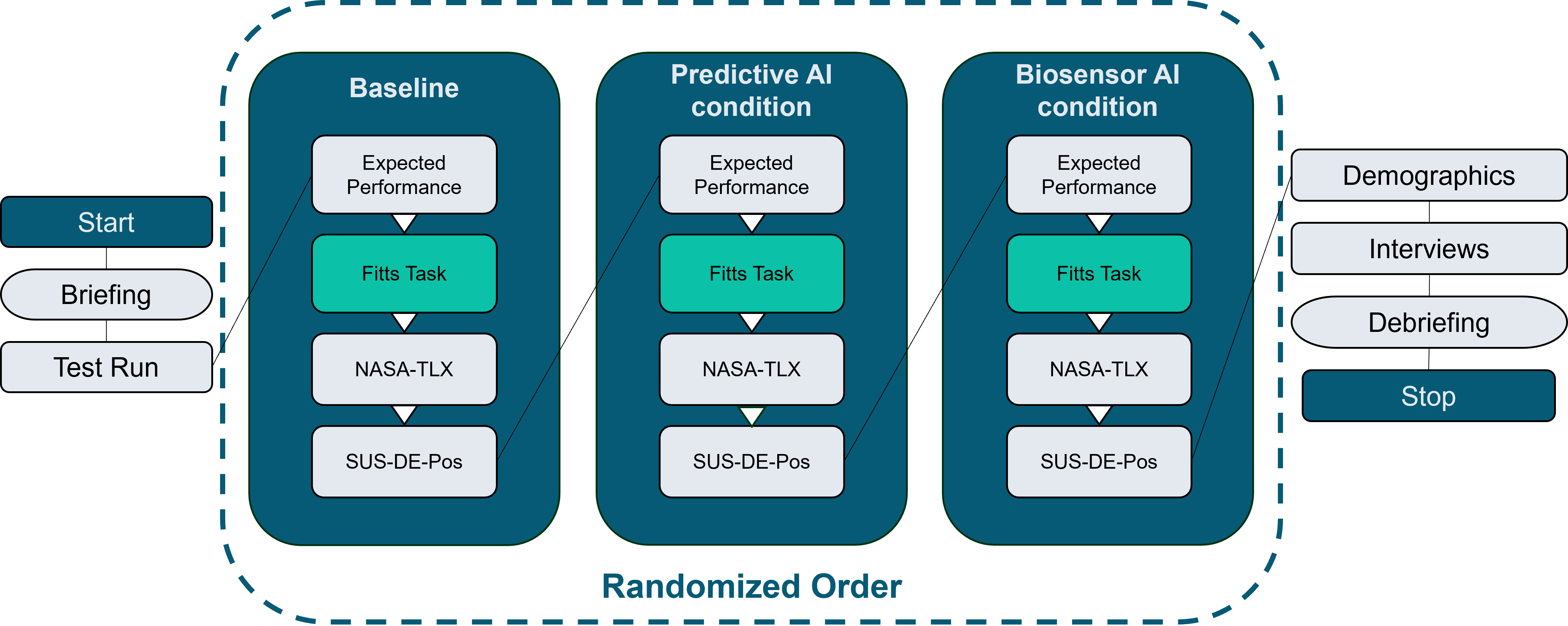}
\caption{Visualisation of the study procedure. The process started with a briefing explaining the conditions and tasks. Then participants completed a short test run, before proceeding to complete one of three experimental blocks. Each block has the same sequence of four tasks: Expected Performance, Fitts' Law Task, NASA-TLX, and SUS-DE-Pos. After completing all blocks, the procedure continued with Demographics, Interviews and a final debriefing.}
\Description{Flowchart showing the sequence of study steps. The process begins with Start, followed by Briefing and a test run. Participants then complete one of three experimental blocks: BL, PAIC, or BIOAIC. Each block has the same sequence of four tasks: Expected Performance, Fitts' Law Task, NASA-TLX, and SUS-DE-Pos. After completing all blocks, the procedure continues with Demographics, Interviews, Debriefing, and ends with Stop.}
\label{fig:ablauf}
\end{center}
\end{figure}

To investigate how AI washing shapes user expectations and performance, we conducted a controlled study using a standardised Fitts' Law task with a supposedly \emph{AI-assisted} computer mouse. When devices such as AI-branded mice influence users’ expectations without providing genuine functional support, this may produce placebo-like effects if these expectations shape post-task subjective assessments or behaviours~\citep{kosch_placebo_2022}. This highlights the need for systematic, standardised evaluation methods to identify and measure such effects, ensuring that claims of \emph{AI assistance} are supported by more than marketing rhetoric. Building on this perspective, our study investigates the impact of labelling a familiar input device as “AI-assisted” on performance expectancy, subjective perceptions, and objective performance. We use validated HCI scales to assess subjective perceptions, and Fitts' Law as the experimental task because it is one of the most established and widely validated models for evaluating input performance. Its robustness and applicability to the computer mouse make it well-suited for examining how supposed AI support may affect objective performance. 

\subsection{Methodology} \label{sec:Methodology}

\subsubsection{Task} 

Our study employed a standard Fitts' Law pointing task to measure the speed–accuracy trade-off in mouse-based input. Originally introduced in 1954~\citep{Fitts1954}, Fitts' Law models movement time in pointing tasks as a function of target distance and size, thereby capturing the relationship between speed and accuracy. Grounded in information-processing theory and Shannon’s 17th theorem, it has been extensively validated and remains the standard framework for evaluating input devices and interaction techniques~\citep{soukoreff2004, norman_fitts_2018, mackenzie1992fitts}. Subsequent refinements have improved model fit by incorporating measures such as effective target width and error rate, offering more precise yet interpretable accounts of human performance~\citep{douglas1999iso, Kong2006}.

\subsubsection{Conditions} \label{subsubsec:Conditions}

\begin{figure}[t!]
    \centering
    \includegraphics[width=1\linewidth]{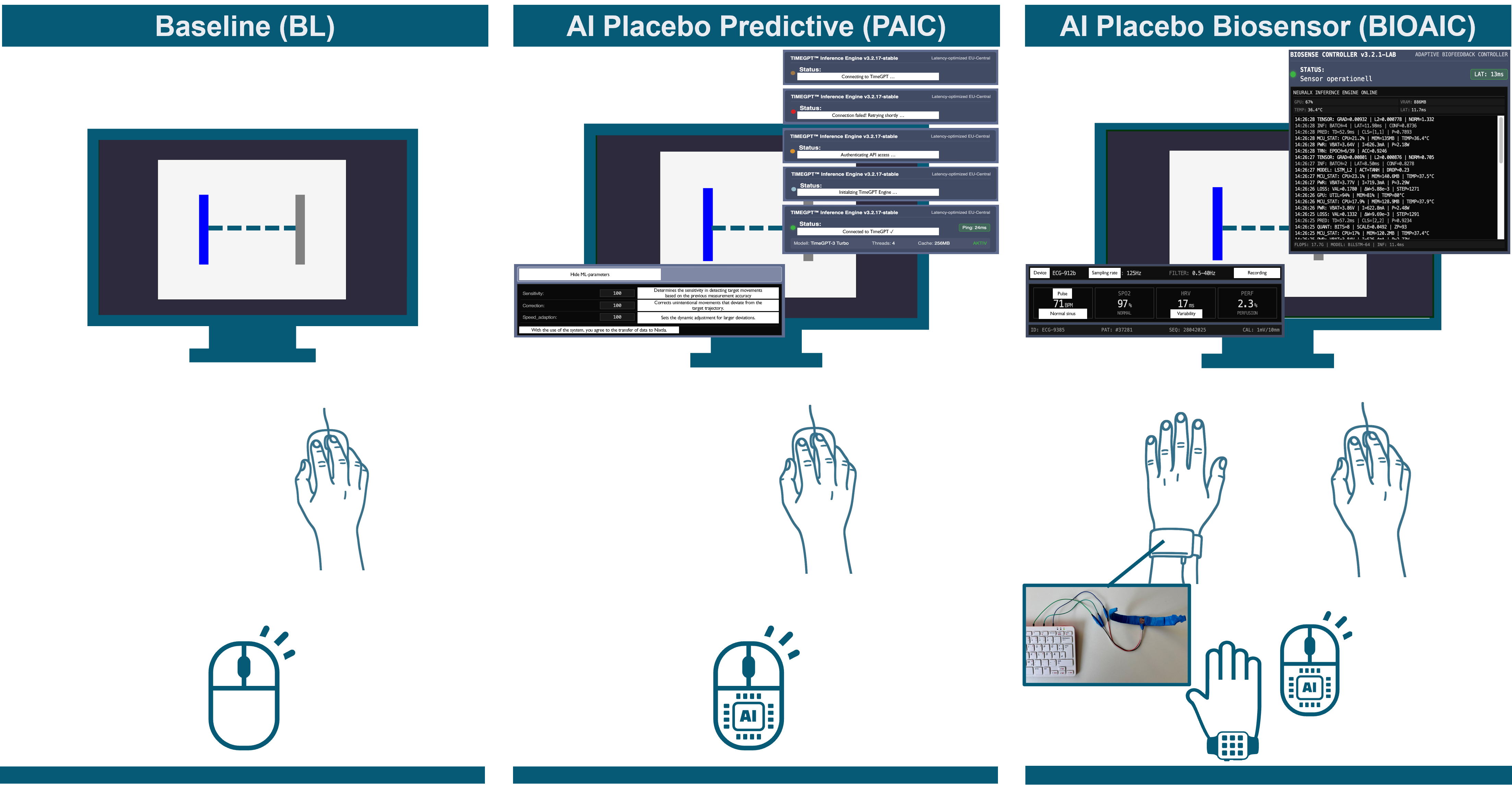}
    \caption{Illustration of the Fitts' Law clicking task and the different interface and hardware elements we used to achieve the deceptions of the experimental tasks. Text in white boxes shows the translations of the German versions that the participants saw.}
     \Description{Interfaces and setup for baseline, predictive AI, and biosignal-enhanced AI conditions.
The figure is divided into three panels. Each panel shows a computer display and a participant’s hand interacting with a mouse, corresponding to one condition.
BL: The screen displays a one-dimensional Fitts' Law task with a blue starting bar, a dashed line, and a grey target bar. A hand illustration below shows mouse input.
PAIC: The same task is shown on the display, overlaid with windows representing a loading screen and a dropdown with AI system parameters. A single hand with a mouse is illustrated below. BIOAIC: The display again shows the task, with additional biosignal interface windows reporting heart rate, oxygen saturation, and related values. Below the display, two hand illustrations indicate the use of both the mouse and wearable sensors. An additional photo shows the biosignal hardware: Raspberry Pi with connected sensor, strap, and connected cables.}
    \label{fig:interfaces}
\end{figure}

Participants completed the same Fitts' Law task under three conditions. In the baseline condition (BL), a standard Hewlett-Packard (HP) mouse was used without modifications. In the second condition, referred to as the predictive AI condition (PAIC), participants were deceived into believing that the HP mouse was “enhanced with predictive AI” based on machine learning. In this condition, a pop-up window in the test software informed participants that cursor control would be \emph{supported by machine learning, which corrected jerky or inaccurate movements and accelerated longer cursor trajectories. The system processed mouse input by analysing (x, y) coordinates and velocity vectors in a 250 ms sliding window with 50 ms overlap. Using the transformer-based model \textit{TimeGPT}, it predicted the optimal target speed for the subsequent 100 ms and adjusted parameters during execution}. In PAIC, a connection to the \textit{TimeGPT} API provided by Nixtla was simulated~\citep{nixtla_timegpt_2024}. To enhance credibility, participants signed a consent form at the start of the study, and a minor setup error was staged during connection to mimic a realistic environment. A drop-down menu displayed system parameters, which were initially set to a supposed default value of 100\%. 
In the third condition, referred to as the biosensor condition (BIOAIC), participants were deceived into believing that the mouse was “enhanced by a biosensor using machine learning” to optimise cursor movements in real time based on physiological signals. Before beginning the BIOAIC condition, a pop-up window informed the participants that \emph{the system would be measuring heart rate, oxygen saturation, and perspiration levels using a bidirectional LSTM model, called \textit{PhysioBERT}, which would enable mouse control based on biosignal input.} To create the illusion of physiological support, participants were asked to rest their left hand on a sensor. To prevent “measurement errors”, participants wearing wristwatches were asked to remove them before starting the task. The sensor was attached to a Raspberry Pi 400 via a wristband, which appeared to monitor hand position throughout the task. In reality, no measurements were taken; the console display was purely simulated. This setup offered convincing visual feedback that mimicked the logic of “real-time biosignal-enhanced AI support" without any actual data collection or system influence. The BIOAIC condition was designed following~\citet{kosch_placebo_2022} and rests on the same underlying logic: users’ mental models of a system can shape how its performance is expected and evaluated~\citep{garciagarciaSeeingMovementBelieving2021}. This motivated the specific research question of whether the elaborateness of an AI-related framing modulates its expectation effect and subsequent performance ratings. Many companies market biosensing hardware together with AI-related claims, often implying a more personalised form of assistance, for example, RingConn markets its smart ring together with the promise of “your personalized AI partner”~\citep{ringconn_website}. The comparison between BIOAIC and PAIC therefore enables an assessment of whether a more physiologically grounded AI narrative increases expectations beyond those associated with a software-only AI claim. This question has direct practical relevance for understanding how AI-related claims are incorporated into the marketing of consumer hardware products. 
We deliberately did not include a condition in which the mouse provided genuine AI-enabled assistance. In our study, AI washing refers specifically to labelling a system as AI-assisted while its underlying functionality remains unchanged. In a within-subjects design, an active AI condition would make it difficult to isolate the expectation effect attributable to the label alone from real performance benefits and could introduce carry-over across later blocks. Our goal was to measure what the label does in the absence of functional change.

An overview of the user interfaces of all three conditions in the test software can be seen in \autoref{fig:interfaces}. Data processing was presented as running on Amazon SageMaker, reflecting typical features of modern AI infrastructures~\citep{liberty2020elastic}. 

\subsubsection{Procedure} \label{subsubsec:Procedure}

The study employed a within-subjects design, which increases statistical power by reducing between-subject variability and has been widely used in experiments on AI placebos \citep{kosch_placebo_2022, denisova2015placebo}. Each participant was randomly assigned to one of six possible sequences (see ~\autoref{fig:ablauf}). Participants were tested individually in the same lab room under a standardised setup using identical hardware with unchanged settings across all sessions. A standard office setup was chosen to avoid demand effects from specialised equipment. Participants interacted with the tasks using a wired HP office mouse (\textit{HSA-D004M}) on a Dell ultrawide monitor, with the experiment run on a MacBook Air (M1, 2020) with 8 GB RAM, ensuring consistent display resolution and processing performance across trials. After greetings, participants were briefed that they would complete a series of clicking tasks under three different experimental conditions, and would answer questions before and after each task. They then received an overview of the three conditions and what distinguished them. Next, the Fitts' Law task was explained asking participants to complete a brief unrecorded practice run to familiarise them with the task before starting the experiment (three difficulty levels, five clicks each). Before the test run, a pop-up window simulated standard authentication steps (\textit{Verify access data}, \textit{Initialise ML resources}) to further make the deception believable. Following the practice run, participants received a summary of the experiment procedure and the different mouse types. Before the start of a new condition, participants were presented with a technically worded description of the current condition through a pop-up window, which was also read aloud by the study supervisor. They then proceeded to the Fitts' Law Task, where each of the three conditions comprised the same 15 (\(W\),\(D\)) combinations in random order. To ensure sufficient data, 15 clicks were performed for each of these combinations following~\citet{soukoreff2004}. A total of 675 clicks were recorded per participant. The specific values of the (\(W\),\(D\)) combinations can be found in the ~\autoref{sec:appendix}. They cover a variability of \(ID\)= 6.1 bits, consistent with the recommended range of 2--8 bits between (\(W\),\(D\)) combinations~\citep{soukoreff2004}. After completion of all three conditions, participants provided demographic information, before being interviewed, and then debriefed on the study goals and deceptions. 

\subsection{Hypotheses}

Building on the theoretical concepts and considerations outlined above, we formulate hypotheses that distinguish between expectancy effects, subjective performance assessments (perceived usability and perceived workload) and objective performance measures (error rate, model fit, and throughput). 

\begin{itemize} 
    \item \textbf{Subjective Performance} 
        \begin{itemize} 
            \item \textbf{H1a:} Before completing the task, participants anticipate higher performance in the predictive AI condition compared to the baseline condition.
            \item \textbf{H1b:} Before completing the task, participants anticipate higher performance in the biosensor condition compared to the baseline condition.
            \item \textbf{H2:} Perceived overall workload (NASA-TLX) will differ between the three experimental conditions.
            \item \textbf{H3:} Perceived performance (NASA-TLX subscale) will differ across the three experimental conditions. 
            \item \textbf{H4:} Perceived usability (System Usability Scale) will vary between the three experimental conditions.
        \end{itemize}
    \item \textbf{Objective Performance} 
        \begin{itemize} 
            \item \textbf{H5:} There is no significant difference in error rate between the three experimental conditions.
            \item \textbf{H6:} There is no significant difference in model fit ($R^2$ of Fitts' Law regression) across the three experimental conditions.
            \item \textbf{H7:} There is no significant difference in throughput (bits per second; a combined index of pointing speed and accuracy from Fitts' Law) between the three experimental conditions.
        \end{itemize} 
\end{itemize}

\subsection{Participants}
To determine the required sample size of \(N = 28\), we conducted a priori power analysis in G$^{*}$Power 3.1~\citep{faul2007gpower} with a repeated-measures ANOVA (within factors: 1 group, 3 measurements). We used conventional values (\(\alpha = .05\), 80\% power) to detect a medium effect-size (\(f = .25\)) and default repeated-measures parameters (\(r = .50\), \(\epsilon = 1.0\))~\citep{faul2007gpower, faul2009statistical}. We considered this medium effect size a conservative and reasonable target, as directly comparable prior work in mouse-based interactions is limited. Participants were students aged 19–32 years (\(M = 22.86\), \(SD = 2.85\)) and were primarily recruited through snowball sampling, with the sample largely drawn from STEM-adjacent disciplines (96.4\%). Nineteen (67.9\%) identified as male and nine (32.1\%) as female. Gender was assessed using an inclusive set of options (male, female, non-binary, no answer, free-text), following recommendations for gender-sensitive survey design in HCI~\citep{spiel2019gender}. To avoid stereotype activation effects on performance~\citep{shih1999stereotype}, demographic information was collected only after completing the task. To increase the credibility of the adaptive mouse settings and strengthen the placebo manipulation, participants were additionally asked to confirm that they had no cardiovascular conditions. One person was excluded based on this criterion. 

\subsection{Measures}

\subsubsection{Subjective Performance Assessment}
The expected performance was assessed before interaction using an item adapted into German from~\citet{kosch_placebo_2022} measured on a seven-point Likert scale\footnote{Note that~\citet{kosch_placebo_2022} also used UTAUT measures to assess performance, which we did not adopt for this study.}. Before using the mouse, participants received the following instructions:

\begin{quote}
\textit{“You are now testing the [...] condition. How do you think you will perform compared to the other configurations?”}
\end{quote}

The response scale ranged from 1 (= worse) to 7 (= better) and was intended to reflect subjective expectations regarding one's own performance in comparison to the other mouse types.

To assess subjective experiences for the mouse-based interaction task, we used versions of two of the most established and widely used HCI instruments~\citep{perrigMeasurementPracticesUser2024}: the NASA-TLX~\citep{sandra2006} and System Usability Scale~\citep{brooke_sus_1996}. 

\subsubsection{NASA-TLX} The NASA-TLX was employed to quantify perceived workload across six dimensions: mental demand, physical demand, temporal demand, performance, effort, and frustration~\citep{hart1988nasa}. This is particularly suitable because it allows us to assess whether identical pointing tasks and hardware are experienced as more or less effortful depending on the described experimental condition. To keep the study duration manageable, we used the unweighted Raw TLX, which is widely adopted and has been shown to retain sensitivity comparable to the weighted version. We relied on a publicly available German translation~\citep{nasatlx_de}, whose psychometric evaluation demonstrated a coherent one-factor structure, satisfactory explained variance ($R^2$= 56.9\%) and high internal consistency ($\alpha = .84$), though confirmatory fit indices were less conclusive~\citep{flaegelNationalAeronauticsSpace2019}. The internal consistency was just below acceptable to good when considering Cronbach’s $\alpha$ ranging from .692 to .704; however, the more suitable indicator \emph{McDonald’s $\omega$} ranged from .781 to .803 within conditions. Averaging across conditions, we achieved Cronbach’s $\alpha$ of .693, and $\omega$ of 0.781, indicating that the NASA-TLX provides a reasonably reliable measure of subjective workload. 

\subsubsection{SUS-DE-Pos} 
To assess perceived usability---how intuitive and easy the system is to operate---we used the SUS-DE-Pos~\citep{perrig2024sus}. This version builds on Rummel’s German System Usability Scale~\citep{rummel2016sus}, reformulating all negative items into positive wording. We employed it to evaluate whether the same physical mouse interaction is experienced as more or less usable depending on the described experimental condition. Its development included language expert review and preregistered psychometric testing in two large-scale studies ($N > 1100$).

The SUS-DE-Pos showed high internal consistency (Cronbach’s $\alpha$ and McDonald’s $\omega$ $\geq$ .90), strong convergent validity with pragmatic usability measures, and weaker correlations with hedonic quality constructs, supporting divergent validity. Confirmatory factor analyses supported a unidimensional structure, and strict measurement invariance was demonstrated across desktop and mobile contexts. Positive-only wording also reduced participant response errors. In our sample, reliability was consistently high, with Cronbach’s $\alpha$ ranging from .891 to .909 and McDonald’s $\omega$ from .915 to .929 within conditions. Averaged across conditions, we achieve excellent internal consistency with Cronbach’s $\alpha$= 0.90 and McDonald’s $\omega$= 0.92.

The SUS-DE-Pos thus offers a validated and reliable measure of usability in German that remains compatible with the original SUS scoring procedure. Items were answered on a five-point Likert scale from 1 (\textit{‘Strongly disagree’}) to 5 (\textit{‘Strongly agree’}), administered in analogue form after each condition. Full study materials are available on the OSF repository. Taken together, our analyses provide evidence that the instruments are suitable for our context and improve on the use of ad-hoc measures still common in HCI \citep{perrigMeasurementPracticesUser2024}. 

To assess demand effects, we followed~\citet{denisova2015placebo} and asked additional open-ended questions to check whether participants perceived the manipulation as credible. A qualitative analysis of these comments was carried out to record subjective attributions regarding the functionality of the supposedly supported conditions.

\subsubsection{Objective Performance Assessment}

To assess objective performance, we applied Fitts' Law. \citeauthor{Fitts1954} introduced the \emph{Index of Difficulty} ($ID = \log_2(2D/W)$), where $D$ is movement distance and $W$ target width. Throughput ($TP$) is defined as $TP = ID/MT$, with $MT$ denoting mean movement time. The relationship between $MT$ and $ID$ was analysed using linear regression~\citep{soukoreff2004}, with the slope reflecting sensitivity to task difficulty and $R^2$ indicating model fit.  

\citet{norman_fitts_2018} recommend using the effective width $W_e = 4.133 \cdot SD$, where $SD$ is the standard deviation of hit points, computed here with the combined coordinate system method ($CC$)~\citep{Kong2006}. This yields the effective index of difficulty $ID_e$, which improves regression quality~\citep{mackenzie1992fitts}. Throughput was then calculated as $TP = \tfrac{1}{n}\sum_{i=1}^n (ID_{e_i}/MT_i)$.  

Error rate was included as an additional measure~\citep{soukoreff2004}, following~\citet{burno2015fitts}: $ER = \tfrac{1}{n}\sum_{i=1}^n e_i \cdot 100\%$, where 
$e_i = 1$ if click $i$ was incorrect and $e_i = 0$ otherwise. Here, $n$ is the number of clicks per participant and mouse type.  

In summary, the measured objective variables were regression quality ($R^2$), throughput ($TP$), a combined measure of speed and accuracy that reflects information transmitted per second, and error rate ($ER$), the proportion of incorrect clicks.  

\subsection{Software Implementation}

To support reliable data capture and a credible simulation of placebo conditions, we developed a custom web application with a reactive front end and a dedicated back end\footnote{The complete implementation is available in the GitHub repository: \url{https://github.com/luisaellamueller/fittslawplacebo}}.
The front end, built in React as a single-page application, enabled precise control over task presentation and timing, reducing latency that could otherwise affect performance measurements. It communicated with the back end via a REST API, ensuring clear separation between the experimental logic and data storage. 
The back end, implemented with Node.js and Express.js, managed all experiment logic, validated incoming data, and stored it in a structured MySQL database. This design ensured that each dataset was correctly linked to participants and experimental conditions, preserving the integrity of the analysis and supporting reproducibility.

\section{Results} \label{chap:results}

We first analyse participants’ subjective assessments, followed by objective performance. Unless noted otherwise, preregistered assumptions were met. For sphericity violations, Greenhouse–Geisser corrections were applied. Normality of residuals was checked via histograms, QQ-plots, and the Shapiro-Wilk test. When omnibus tests were significant, post-hoc comparisons with Bonferroni correction addressed multiple testing.

\subsection{Subjective Performance}

\begin{figure}[h!]
  \includegraphics[width=\textwidth]{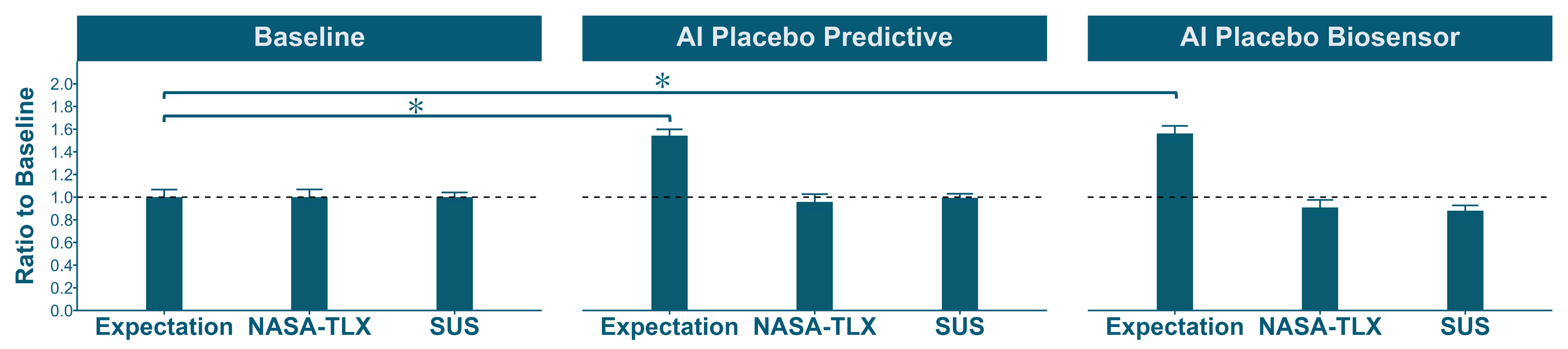}
  \caption{Overview of subjective measure experimental outcomes. Although participants expected higher performance with AI, no differences appeared in post-task subjective measures. Barplot values are normalised relative to baseline (set to 1). $^{*} p < .001$ (Bonferroni-corrected).}
\Description{Results for three conditions: baseline, predictive AI support, and biosignal-enhanced AI support.
This bar chart compares expected performance, NASA-TLX, and SUS-DE-Pos ratings. Only the expected performance differs across conditions, with higher values in both placebo conditions.}
  \label{fig:teaser_small}
\end{figure}

\subsubsection{Expected Performance}

The average expected performance value in PAIC and in BIOAIC was significantly higher than in the baseline condition;~\autoref{tab:subjective_measures} summarises these results. The RM-ANOVA revealed a significant main effect of condition, $F(2, 54) = 23.30$, $p < .001$, $\eta^2 = 0.463$. Since the data in all three conditions deviated from a normal distribution, we conducted non-parametric tests to check the robustness of the results. A Friedman test, suitable for repeated measures with non-normal data~\citep{corder2009friedman}, also showed a significant difference between conditions, $\chi^2(2) = 23.53$, $p < .001$.

Both parametric and non-parametric tests revealed significantly higher expectations for PAIC compared to baseline, and between BIOAIC and baseline condition (supporting H1a and H1b). 
However, no significant differences between the experimental conditions were found (see ~\autoref{tab:param_wilcoxon}). 

\begin{table}[h!]
\centering
\begin{tabular}{lccc ccc}
\toprule
& \multicolumn{3}{c}{\textbf{$t$-Tests}} & \multicolumn{3}{c}{\textbf{Wilcoxon-Tests}} \\
\cmidrule(lr){2-4} \cmidrule(lr){5-7}
\textbf{Comparison} & $t$ & $p$ (adj.) & $\eta^2_p$ & $T$ & $p$ (adj.) & $r$ \\
\midrule
PAIC vs. BIOAIC & -0.26 & 1.000 & < .001 & 88.00 & 1.000 & 0.04 \\
PAIC vs. BL  & 5.43  & < .001* & 0.42 & 31.00 & < .001* & 0.51 \\
BIOAIC vs. BL  & 5.51  & < .001* & 0.39 & 22.00 & < .001* & 0.49 \\
\bottomrule
\end{tabular}
\caption{Comparison of expected performance between conditions. Parametric t-tests (Bonferroni-corrected) and Wilcoxon signed-rank tests for the same contrasts showed significantly higher expected performance in both AI-labelled conditions relative to baseline, with no significant difference between the two AI-labelled conditions.}
\label{tab:param_wilcoxon}
\end{table}

\subsubsection{NASA-TLX}

For interpretative context, raw TLX scores range from 0 to 100, with higher values indicating greater perceived workload; values between 30 and 50 are commonly interpreted as moderate to low workload~\citep{hart1988nasa}. 
Descriptively, BL showed the highest subjective workload on average, while PAIC and BIOAIC were slightly lower, BIOAIC being lowest (see ~\autoref{tab:subjective_measures}).
No significant differences in subjective workload were observed across conditions (RM-ANOVA: $F(2,54)=0.79$, $p=.46$, $\eta^2=.03$). Because we were especially interested in performance, we also looked at the performance subscale individually, which showed no significant differences ($F(2,54)=0.16$, $p=.85$, $\eta^2=.01$), failing to support H2 and H3. This means that no influence of condition on subjective workload could be supported.

\subsubsection{SUS-DE-Pos}

SUS-DE-Pos scores were converted to the standard 0 to 100 SUS scale, ranging from 0 (lowest perceived usability) to 100 (highest perceived usability). We used standard conversion to obtain a composite score: subtracting 1 from each raw item score, summing the recoded item scores, and multiplying the total by 2.5. According to the adjective rating scale by~\citet{bangor2009adjective}, scores above 85 are classified as Excellent, and scores between 70 and 85 as Good. The observed scores, BL 83.2, PAIC 82.7, BIOAIC 73.3, therefore indicate good to near excellent perceived usability across conditions (see ~\autoref{tab:subjective_measures}). Although RM-ANOVA indicated a significant effect ($F(2,54)=4.04$, $p=.02$, $\eta^2=.13$), violations of normality required non-parametric testing. We subsequently performed a Friedman test, which showed no significant differences ($\chi^2(2)=4.69$, $p=.10$). Post-hoc Wilcoxon tests with Bonferroni correction likewise revealed no significant contrasts (PAIC vs. BIOAIC: $p_\text{adj}=.10$; PAIC vs. BL: $p_\text{adj}=1.00$; BIOAIC vs. BL: $p_\text{adj}=.18$). Although BL and PAIC showed a descriptive advantage over BIOAIC, no statistically significant differences were observed. Hypothesis H4 is therefore not supported.

\begin{table}[h!]
\centering
\begin{tabular}{lccc ccc ccc}
\toprule
 & \multicolumn{3}{c}{\textbf{Expected Performance}} & \multicolumn{3}{c}{\textbf{NASA-TLX}} & \multicolumn{3}{c}{\textbf{SUS-DE-Pos}} \\
\cmidrule(lr){2-4} \cmidrule(lr){5-7} \cmidrule(lr){8-10}
\textbf{Condition} & \textbf{M} & \textbf{SD} & \textbf{SE} & \textbf{M} & \textbf{SD} & \textbf{SE} & \textbf{M} & \textbf{SD} & \textbf{SE} \\
\midrule
BL & 3.61 & 1.29 & 0.24 & 39.23 & 14.26 & 2.69 & 83.21 & 18.19 & 3.44 \\
PAIC     & 5.57 & 1.03 & 0.20 & 37.59 & 14.15 & 2.67 & 82.68 & 15.94 & 3.01 \\
BIOAIC   & 5.64 & 1.28 & 0.24 & 35.68 & 13.70 & 2.59 & 73.30 & 20.51 & 3.88 \\
\bottomrule
\end{tabular}
\caption{Descriptive statistics of subjective measures across conditions. Expected Performance was rated on a 1--7 scale (higher = higher expected performance). NASA-TLX values are scores on a 0--100 scale (higher = greater perceived workload; values indicate low-to-moderate workload, approximately in the 30th--40th percentile \citep{hertzumReferenceValuesSubscale2021}). SUS-DE-Pos values use standard scoring on a 0--100 scale (higher = greater perceived usability; scores between 70 and 85 are commonly interpreted as good \citep{bangor2009adjective}).}
\label{tab:subjective_measures}
\end{table}

\subsubsection{Qualitative Evaluation of Placebo Conditions} 

\textbf{Perceived Smoothness and Efficiency in PAIC}: Across both placebo conditions, most participants reported noticing some form of AI support and generally evaluated it positively. PAIC was consistently associated with greater ease and speed, described as “smoother,” “faster,” and “more intuitive”. One participant noted, “I found the first setting (PAIC) very pleasant, as I didn’t have to make such large hand movements” (P2). The accelerated mouse movement was frequently reported as beneficial (P4, P10), though a few participants occasionally reported overshooting (P1, P7, P25).

\textbf{Mixed Experiences with BIOAIC}: Perceptions of BIOAIC were more ambivalent. Some participants valued its “efficiency” (P21) and “intuitiveness” (P16, P18), while others critiqued aspects of the physical interaction. For example, P11 described needing to “concentrate on my hand,” and P20 expressed uncertainty: “I didn’t know exactly where the feedback was integrated.”

\textbf{Uncertainty About the Nature of Support}: Several participants speculated whether the perceived assistance stemmed from practice effects rather than genuine system intelligence. As P15 reflected, “Maybe I was just more practised.” Importantly, even when directly prompted with a question about deception at the end of interviews, no participant suspected deception or questioned the study’s integrity.

Taken together, PAIC reliably evoked perceptions of smoothness and efficiency, while BIOAIC provoked more varied interpretations, highlighting how the physical and narrative design of interaction mediates perceptions. Across both conditions, participants attributed differences in performance to supposed AI features, underlining how expectation effects may not manifest in self-report ratings but in broader experiences of interaction.

\subsubsection{Correlation Between Expected Performance and Throughput}
Following \citet{kosch_placebo_2022}, we conducted an exploratory Spearman’s rank correlation between subjective and objective performance. In the PAIC, average throughput and expected performance showed a weak negative but non-significant correlation ($r_s = -0.22$, $p = 0.25$). The BIOAIC condition revealed a negligible positive correlation ($r_s = 0.02$, $p = 0.91$), and the baseline condition a negligible negative one ($r_s = -0.02$, $p = 0.93$). Thus, no significant association between expected performance and throughput was observed.

\subsection{Objective Performance} \label{objective_results}
\label{sec:objektive_ergebnisse}
\begin{figure}[t]
  \includegraphics[width=\textwidth]{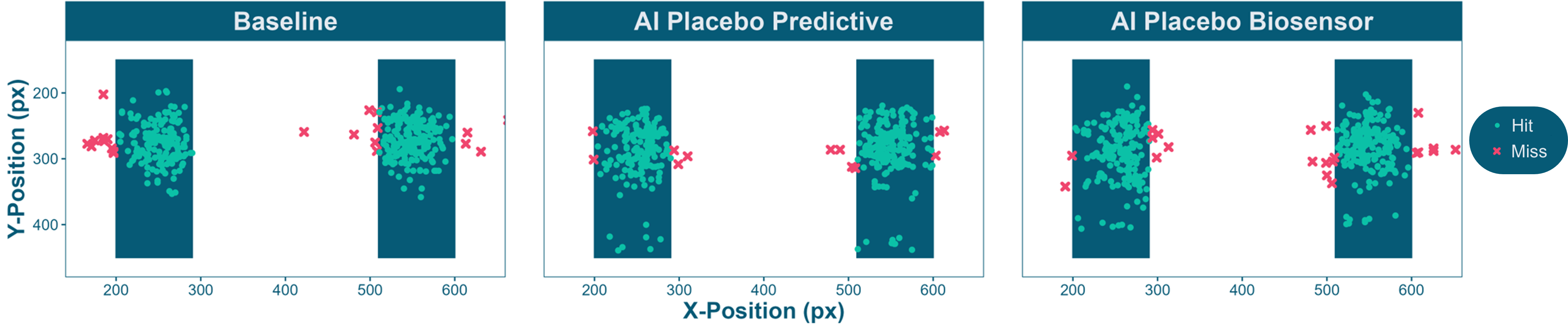}
  \caption{Visualisation of the one-dimensional Fitts' Law task in which participants alternated clicks between left and right targets under three conditions. Hits are shown in teal and misses in red. Each condition produced similar error patterns. The horizontal spread across conditions is similar, but vertical dispersion differs along the y-axis, without clear differences between conditions.}
\Description{This graph shows click distributions in a one-dimensional Fitts' Law task in which participants alternated between left- and right-target clicks. Six vertical bars indicate target positions at a distance D = 550. Both correct and incorrect clicks are shown. Horizontal spread across conditions is similar, but vertical dispersion differs along the y-axis, without clear differences between conditions.}
  \label{fig:errors_fittstask}
\end{figure}

\begin{table}[h!]
\centering
\label{tab:my-table}
\begin{tabular}{lcccccccc}
\toprule
 & \multicolumn{3}{c}{\textbf{Throughput}} & 
   \multicolumn{3}{c}{\textbf{Error Rate}} & 
   \multicolumn{2}{c}{\textbf{Model Fit}} \\
\cmidrule(lr){2-4} \cmidrule(lr){5-7} \cmidrule(lr){8-9}
Condition & Bits/s & SD & SE & Percentage & SD & SE & Equation & $R^2$ \\
\midrule
BL     & 5.54 & 0.69 & 0.13 & 9.00\% & 3.27 & 0.63 &  $MT = 0.1998 \cdot ID - 0.0231$ & 0.936  \\
PAIC   & 5.57 & 0.66 & 0.12 & 8.21\%  & 3.44 & 0.66 & $MT = 0.1923 \cdot ID - 0.0016$ & 0.947 \\
BIOAIC & 5.60 & 0.69 & 0.13 & 8.97\%  & 2.75 & 0.53 & $MT = 0.1971 \cdot ID - 0.0246$ & 0.952  \\
\bottomrule
\end{tabular}
\caption{Descriptive statistics of objective measures across conditions. Objective performance was largely comparable across all three conditions, with only small differences in throughput, error rate, and model fit.}
\label{tab:objective-measures}
\end{table}

\subsubsection{Error Rate}

Following the recommendation of~\citet{soukoreff2004}, subjects whose error rate was more than three standard deviations above or below the mean were excluded. One person had to be excluded from the analysis because their error rate was more than three standard deviations above the mean (error rate > 19.58\%). No peculiarities in the patterns of errors could be seen between conditions (see \autoref{fig:errors_fittstask}). The mean error rates differed slightly between the conditions, as can be seen in~\autoref{fig:errorratemouse}. All conditions showed similar average error rates, with BL and BIOAIC being practically identical and PAIC slightly lower. The descriptive parameters of the error rates per condition are shown in ~\autoref{tab:objective-measures}. The subsequently performed RM-ANOVA revealed no significant difference in the error rate between the three conditions ($F(2, 52) = 1.25$, $p = 0.29$, $\eta^2 = 0.05$). Thus, we failed to reject H5. 

\subsubsection{Model Fit}

\begin{figure}[t]
  \centering
  \begin{minipage}[t]{0.48\linewidth}
    \centering
    \includegraphics[width=\linewidth]{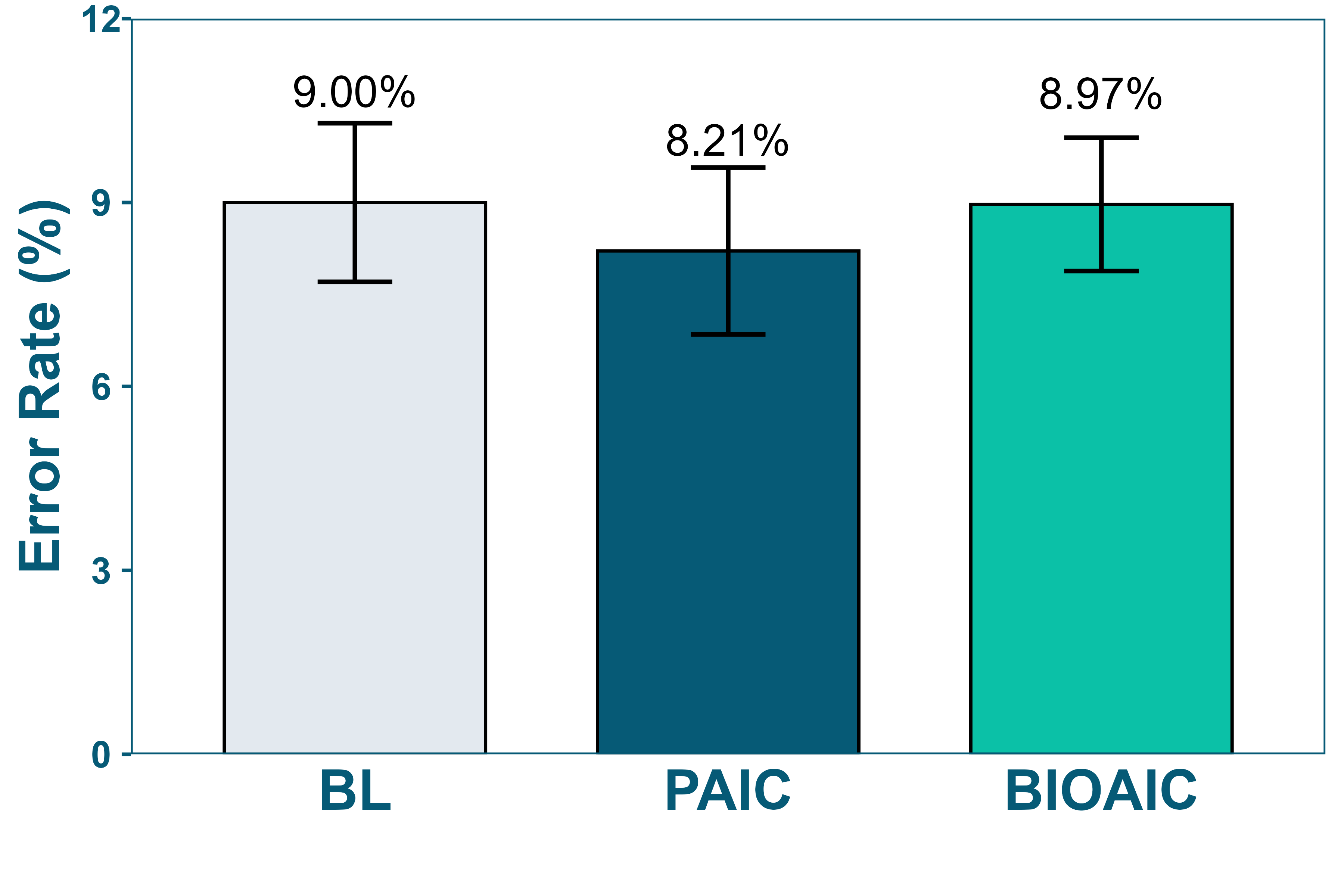}
    \caption{Mean error rate percentage across conditions: Baseline (BL), predictive AI condition (PAIC), biosignal-enhanced AI condition (BIOAIC). Error rates were similar across conditions, indicating no meaningful differences in clicking accuracy between the baseline and AI-labelled mouse settings.}
    \Description{Barplot showing mean error rates with error bars. Baseline has 9.00\%, predictive AI (PAIC) has 8.21\%, and biosignal-enhanced AI (BIOAIC) has 8.97\%. Differences between conditions are small, with PAIC showing the lowest error rate. However, none appear to be significantly different.}
    \label{fig:errorratemouse}
  \end{minipage}%
  \hfill
  \begin{minipage}[t]{0.48\linewidth}
    \centering
    \includegraphics[width=\linewidth]{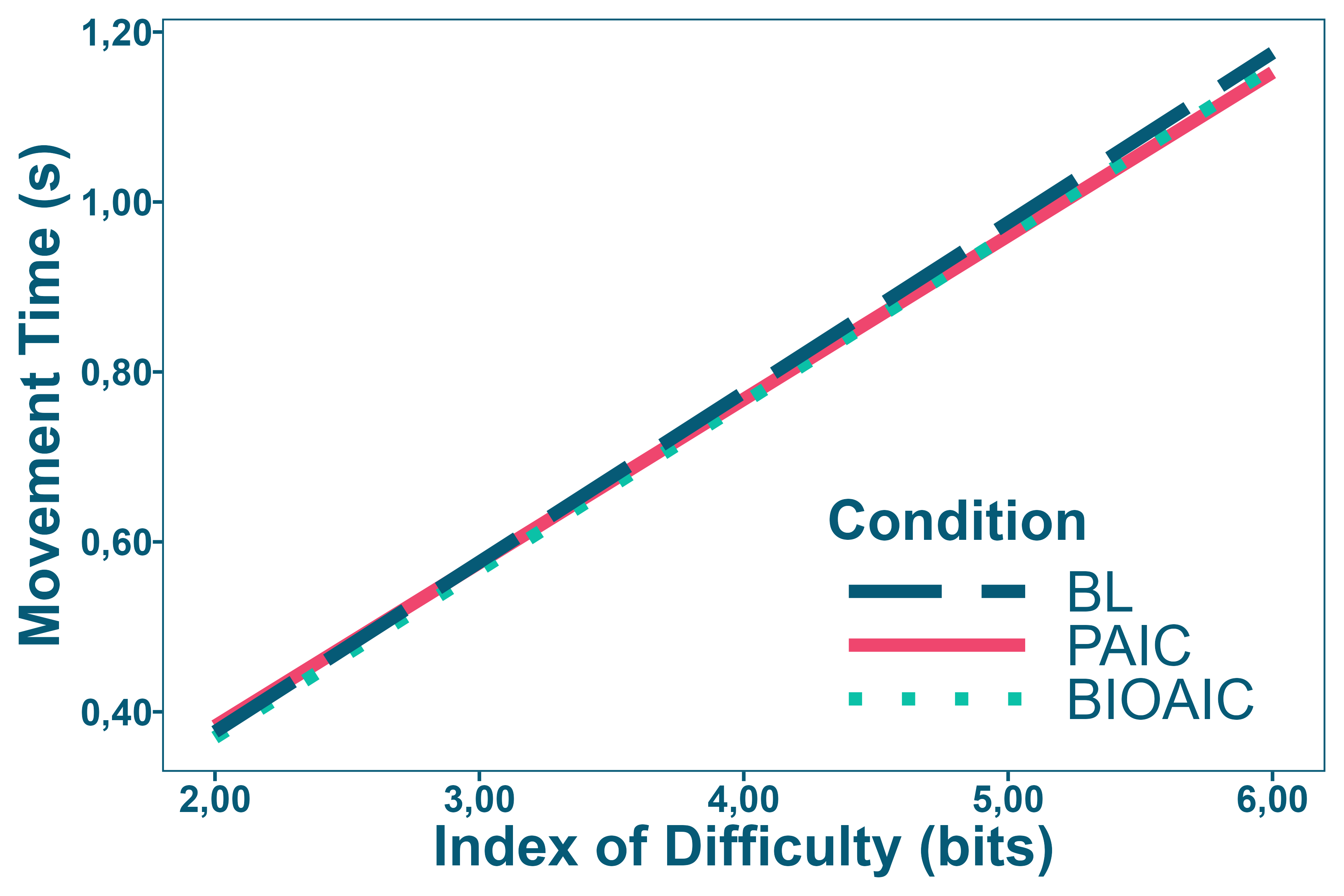}
    \caption{Comparison of Fitts' Law regression models between conditions. Movement time increased with task difficulty in all three conditions; the regression lines overlap closely, showing near-identical slopes and intercepts, indicating no meaningful difference between conditions.}
    \Description{Line graph plotting movement time (y-axis, seconds) against index of difficulty (x-axis, bits). Three regression lines: BL, PAIC, and BIOAIC, overlap closely, showing near-identical slopes and intercepts. This figure indicates no meaningful difference between conditions.}
    \label{fig:three}
  \end{minipage}
\end{figure}

The model quality was evaluated using the coefficient of determination \(R^2\) for the conditions (baseline, PAIC and BIOAIC). According to~\citet{soukoreff2004}, negative intercepts of the regression line up to \(-200\) ms are acceptable in linear regression. All three conditions showed negative intercepts within the explainable range; baseline: \(-20\) ms, PAIC: \(0\) ms, BIOAIC: \(-20\) ms (see \autoref{fig:three} for regression visualisation).
All conditions showed exceptional fit to the data with coefficients of determination \(R^2>0.9\); for detailed regression values see \autoref{tab:objective-measures}.

The coefficient of determination is high and very similar for all mouse types.
The RM-ANOVA showed no significant differences between the conditions ($F(2, 54) = 1.18$, $p = 0.32$, $\eta^2 = 0.04$). These results are consistent with H6.

\subsubsection{Throughput}
The average throughput was calculated for each of the three conditions based on the effective difficulty levels (\(IDe\)). There were no outliers (deviation > 3SD). This means that the data from all 28 participants could be included in the analysis. Across conditions, the highest mean throughput was in BIOAIC, followed by PAIC and then BL. The summary of the descriptive results can be found in ~\autoref{tab:objective-measures}. The RM-ANOVA revealed no significant difference in mean throughput between the three mouse type conditions ($F(2, 54) = 0.38$, $p = 0.68$, $\eta^2 = 0.01$). We therefore did not conduct post-hoc contrast. These results are consistent with H7.

\autoref{tab:hypothesen} summarises the central hypotheses of the study and their empirical support based on the results in ~\autoref{chap:results}. 

\begin{table}[h!]

\centering
\begin{tabular}{p{2cm}p{9.5cm}p{2.5cm}}
\toprule
\textbf{Hypothesis} & \textbf{Content (shortened)} & \textbf{Supported} \\
\midrule
H1a & Expected performance is higher in the PAIC than in the baseline condition. & Yes \\
H1b & Expected performance is higher in the BIOAIC than in the baseline condition. & Yes \\
H2 & Perceived subjective workload differs between conditions. & No \\
H3 & The \textit{“performance”} subscale of the NASA-TLX differs between conditions. & No \\
H4 & Perceived usability (SUS-DE-Pos) differs between conditions. & No \\
H5 & The error rate does not differ significantly between conditions. & Yes \\
H6 & Model fit ($R^2$) does not differ significantly between conditions. & Yes \\
H7 & Throughput does not differ significantly between conditions. & Yes \\\bottomrule
\end{tabular}
\caption[Overview of the hypothesis test results.]{Overview of the hypothesis test results.\\
\footnotesize Note. For directional hypotheses (H1a--H4), “Yes” indicates support for the predicted effect. For preregistered null hypotheses (H5--H7), “Yes” indicates no significant difference, consistent with but \emph{not} confirming the null hypothesis.}\label{tab:hypothesen}
\end{table}

\section{Discussion}

Our study shows that suggested AI support reliably raises expectations of performance but does not translate into measurable improvements in input device assessment. Neither subjective workload, perceived usability, nor any objective metric differed across conditions. This nuance extends the mixed evidence on AI placebos in HCI. While some studies report carry-over effects from expectations to objective performance \citep{kloftAIEnhancesOur2024} or subjective assessments \citep{kosch_placebo_2022}, similar to other studies, we did not find such effects \citep{hsuPlaceboEffectControl2025}.

\subsection{Expectation Effects but no Performance Gains}
These results underscore a methodological concern for AI evaluations. AI-washed interactive technologies can elicit expectation effects without altering actual interaction or subjective assessments. Unlike in medicine, where placebo treatments often produce robust effects, in technology they appear fragile and highly context-dependent~\citep{gonnermann-mullerLetsBeRealistic2025}. Studies in HCI have documented heightened subjective assessments under supposed AI assistance \citep{denisovaPowerUpsDigitalGames2019, kosch_placebo_2022}, but objective performance effects remain inconsistent and often absent. Our findings add to this pattern, suggesting that AI washing for input devices remains limited to expectancy effects, rather than placebo effects in self-reported performance, or objective performance improvements.

This has implications for researchers who compare “regular” technologies with novel ones. In line with earlier calls, we advocate for designs that distinguish genuine system benefits from expectation effects, whether through placebo-control, active-control conditions, or explicit measurement of expectations~\citep{kosch_placebo_2022}. We add to the literature by showing that, even for an everyday technology such as the computer mouse, familiarity did not eliminate heightened expectations induced by AI-labelling. However, familiarity might explain why no subjective assessments improved; descriptively, usability even declined. Thus, AI washing seems effective in inflating expectations for improvement, even for deeply familiar technologies but this did not carry over into actual evaluations in our study, possibly due to interaction familiarity. Interestingly, qualitative feedback suggested perceived improvements despite the absence of measurable gains in standardised scales.

\subsection{Using Fitts' Law to Assess Placebos}
Fitts' Law remains one of the most established methods for evaluating input devices and interaction techniques. It offers a simple yet rigorous benchmark of motor performance, is relatively easy to implement, and allows for comparisons across studies. In our case, throughput, error rate, and model fit proved unaffected by placebo manipulations, although participants qualitatively reported feeling supported by AI. This robustness positions Fitts' Law not only as a benchmark for assessing the performance of new input devices, but also as a valuable control method for objective performance in HCI studies evaluating AI-labelled artefacts. We therefore see Fitts' Law tasks as particularly suited for evaluating novel interaction devices such as game controllers, or other input modalities, as they appear resistant to expectation effects created by promises of novel technology. This underscores the importance of complementing self-reports with objective benchmarks to ensure that claims about AI-supported interaction techniques are not driven by expectation effects alone.

\subsection{Implications for Practice, Accountability, and Regulation}
For practitioners and marketers, our findings offer a cautionary note: simply labelling a product as “AI-assisted” can inflate user expectations without delivering functional benefits. Scholars have argued that the marketing of AI often relies on tactics similar to those used to sell “snake oil”~\citep{narayananAISnakeOil2024}. Our results add an empirical dimension to this concern by showing that the label itself can alter perceptions even when the underlying functionality is unchanged. In domains where safety, accuracy, or efficiency are critical, such inflated expectations may obscure shortcomings and produce immediate expectation-driven harm. This in turn, could undermine trust once systems are scrutinised in real and prolonged use. A hint of this dynamic is evident in our usability ratings: the most elaborate condition (BIOAIC) descriptively fell off from BL and PAIC, dropping from near "excellent" to the “good” category in the SUS-DE-Pos.

For broader implications, we suspect AI washing may not affect all users equally: individuals with limited AI literacy may be less well positioned to critically evaluate marketing claims, raising concerns that inflated expectations could be distributed unevenly across populations. Related work suggests that literacy, prior assumptions, and domain expertise can shape how people interpret AI descriptions and how strongly such descriptions affect their evaluations~\citep{figueiredoPoweredAIExamining2023,gonnermann-mullerLetsBeRealistic2025}. For researchers and policymakers concerned with AI ethics, the issue is therefore not only whether AI systems perform as advertised, but also whether AI labels themselves shape user expectations independent of any underlying functionality. Our findings speak directly to this question by providing evidence that an AI-label alone influences how even familiar systems are perceived.
Our study isolates these effects in a low-stakes motor task without testing downstream harm. However, in safety-critical or decision-support contexts, including clinical tools, workplace automation, and accessibility devices, such misrepresentation could plausibly contribute to misplaced reliance, overconfidence, or failures to seek more appropriate alternatives. The broader concern is also not limited to overly positive framings. Misleading descriptions of AI could, in principle, operate in the opposite direction, discouraging trust or uptake in ways that are not warranted by the system's actual capabilities~\citep{scharowskiTrust2025}. While negative descriptions of sham-AI did not successfully induce such nocebo-like reductions in expectations~\citep{kloftAIEnhancesOur2024}, the heterogeneity of findings in the AI placebo literature leaves open the question of whether representational effects could arise in more than one direction.

Beyond marketing ethics, AI washing raises concerns about informed consent and user autonomy at scale. When users form expectations based on misrepresented capabilities, their ability to make meaningful choices about the tools they adopt may be compromised from the outset. This is particularly consequential in workplace and public-sector contexts, where AI-labelled tools are often procured institutionally rather than chosen by end users, leaving them with limited agency to question or opt out. This echoes critiques about corporate accountability, where companies still strive for quick fixes buying AI just for its label, and leaving the "AI snake oil" market flourishing~\citep{narayananAISnakeOil2024,ludwigAISnakeOil2026}.

Regulatory frameworks concerned with AI transparency, such as the EU AI Act~\citep{eu_ai_act_2024}, address risk, accountability, and transparency obligations for AI systems, but the expectation effects documented here suggest that representational influence can begin even earlier, at the level of product description and marketing claims. Governance mechanisms may therefore need to extend upstream, to cover how AI capabilities are described in product documentation and marketing materials, not only how systems perform once deployed. Consumer protection bodies and procurement guidelines could, for instance, require that AI claims be substantiated by independent benchmarking, so that the gap between marketed and measurable performance is visible to those who rely on these systems.

Taken together, our findings and possible implications highlight both the potential and the boundaries of AI placebo effects. In our study, there was clearly an effect visible at the level of subjective expectations, but it did not extend to perceived workload, perceived usability, or objective performance. At the same time, several methodological limitations of our study need to be acknowledged to properly interpret these results.

\section{Limitations \& Future Work}

This study has several limitations that should be considered when interpreting the results. As is common in within-subjects designs, repetition effects may have occurred. Although we mitigated these through a practice trial and randomisation, learning effects may have improved performance in later trials, while fatigue may have offset these improvements. 
Furthermore, while the use of a standardised Fitts' Law paradigm in a controlled lab setting ensures high internal validity, it limits direct generalisability to more complex or ecologically richer environments. It remains an open question whether the observed expectation effects persist for more sophisticated or unfamiliar technologies. Even though it is known that Fitts' Law is robust and not specific to gender, the sample of $N=28$ was demographically homogeneous, consisting primarily of students, with a gender distribution of 67.9\% men and 32.1\% women. Participants’ mostly STEM-oriented backgrounds may have influenced their prior knowledge of and attitudes towards AI technologies. Although post-task interviews suggested that participants accepted the experimental manipulation, the limited sample size prevented the analysis of subgroup differences. Additionally, we did not measure baseline attitudes towards AI or prior experience with input devices, both of which may have influenced temporal dynamics across blocks. While the within-subjects crossover design reduces confounding from stable individual differences, thereby reducing this concern, these unmeasured factors remain a limitation.

These considerations provide important context for interpreting the findings, but they do not diminish the observed effect on subjective expectations. Future work should investigate the psychological mechanisms underlying the dissociation between expectation and evaluation, including the role of AI literacy and prior experience. It would also be valuable to examine potential nocebo effects among more sceptical users and to disentangle the influence of superficial AI claims from genuinely functional system improvements. It is important to find other validated tasks to extend audits to safety-critical and decision-support contexts, where the gap between expected and actual system capability carries higher stakes, and is not measurable by motor performance. Finally, extending this paradigm to more diverse samples, real-world contexts, and alternative experimental designs using novel input devices and tasks will help determine the robustness and generalisability of our experimental findings.

\section{Conclusion} 

This study provides controlled empirical evidence that AI washing shapes user expectations in the absence of any functional change. Participants anticipated meaningfully better performance when a standard computer mouse was labelled as AI-assisted, yet these expectations produced no corresponding changes in objective performance or post-task self-reports. The dissociation between inflated expectations and measurable outcomes points to a form of user influence that operates independently of system functionality and therefore falls outside the scope of conventional performance audits. This has direct implications for the AI research community. If AI labels systematically alter user beliefs even when system behaviour remains unchanged, transparency requirements applied only to functional capabilities will be insufficient. Users who cannot verify AI claims, whether due to opaque product documentation, limited AI literacy, or the sheer familiarity of everyday devices like the computer mouse, may carry inflated expectations into contexts where accuracy of self-assessment matters. 
Fitts' Law tasks offer one replicable instrument for separating expectation effects from genuine performance gains in empirical audits of AI-labelled artefacts involving motor performance. The absence of measurable changes in subjective or objective interaction outcomes highlights a dissociation between expectations and actual performance. In contrast to other work that reported changes in interaction outcomes, this underscores the context-sensitivity of AI placebo effects. We contribute to the growing literature on AI expectancy and placebo effects by investigating how user beliefs can be shifted through AI washing, without corresponding improvements in subjective attitudes or objective performance using validated instruments. Ultimately, we demonstrate a critical need to extend empirical audits beyond functional capabilities. As AI washing continues to shape user beliefs, evaluating systems must increasingly account for the expectations their labels create, especially in environments where the gap between perceived and actual capability carries high stakes.

\newpage

\section{Generative AI Usage Statement}
For the writing of this manuscript, we used Overleaf’s spell checker, Gemini 3 Pro, and ChatGPT (versions GPT-5.1 and 5.2) exclusively for linguistic polishing and formatting assistance. These tools were not employed to generate original ideas, data, or substantive content. The hand pictogram in \autoref{fig:interfaces} was generated with ChatGPT (GPT-4o).

\section{Ethical Considerations}
Since the study involved human participants and included deliberate deception, we obtained ethics approval from the Ethics Committee of the University of St. Gallen on 15 December 2024 (Reference No.: \textit{HSG-EC20241113}). The deception was non-harmful and necessary to preserve the study’s aims. Participants were informed about the tasks and procedures in advance, except for the manipulation regarding the supposed AI support. After completing the study, participants were fully debriefed about the study’s intent and were given an additional opportunity to withdraw their consent.

\section{Data Availability Statement}

The study was pre-registered on the Open Science Framework (OSF)~\citep{foster2017osf}.\\
Pre-registration: \url{https://doi.org/10.17605/OSF.IO/QZP3D}\\
Study materials: \url{https://osf.io/gvhk3/overview?view_only=6e5957c92ea249ec9b06d2fcb2300542}.
\section{Competing Interests}
The authors declare that they have no known competing financial interests or personal relationships that influence the work reported in this paper.

\section{Acknowledgements} 
We want to thank the study participants for their time and cooperation. We also express our gratitude to the anonymous reviewers for their constructive feedback, which greatly helped improve the manuscript.

\bibliographystyle{ACM-Reference-Format}  
\bibliography{bibliography}      

\newpage
\appendix

\section{Appendix} \label{sec:appendix}
\begin{table}[ht]
\centering
\begin{tabular}{ccc}
\toprule
\textbf{ID (bits)} & \textbf{Width (px)} & \textbf{Distance (px)} \\
\midrule
1.5 & 140 & 300 \\
1.7 & 120 & 350 \\
2.3 & 90  & 400 \\
3.3 & 50  & 450 \\
3.6 & 40  & 450 \\
4.0 & 30  & 450 \\
4.4 & 25  & 500 \\
4.9 & 20  & 550 \\
5.3 & 15  & 550 \\
5.7 & 12  & 600 \\
6.1 & 10  & 650 \\
6.5 & 8   & 700 \\
7.0 & 6   & 750 \\
7.3 & 5   & 750 \\
7.6 & 4   & 780 \\
\bottomrule
\end{tabular}
\caption{Overview of the difficulty levels used.}
\label{tab:difficulties}
\end{table}

\end{document}